\documentclass[journal]{IEEEtran}
\usepackage[T1]{fontenc}
\usepackage[utf8]{inputenc}
\usepackage{cite}
\usepackage{graphicx}
\usepackage{amsmath}
\usepackage{algorithmic}
\usepackage{amssymb}
\usepackage{standalone}
\usepackage{tikz}
\usepackage{amsfonts}
\usepackage{pgfplots}
\usepackage[numbers,sort&compress]{natbib}
\usepackage{subcaption}
\usepackage{array}

\graphicspath{{figures/}}
\newcommand{\abs}[1]{\left|#1\right|}
\newcommand{\norm}[1]{\left|\left|#1\right|\right|}

\newcommand{\card}[1]{\#\{#1\}}
\newcommand{\trans}[0]{\mathsf{T}}

\newcommand{\conj}[0]{\ast}

\newcommand{\bydef}[0]{\triangleq}

\newcommand{\frall}[1]{\forall #1 \in \mathcal{\uppercase{#1}}}

\newcommand{\myth}[0]{-th }
\newcommand{\ch}[1]{{\mathbf{h}}_{#1}}
\newcommand{\ech}[1]{{\bar{\mathbf{h}}}_{#1}}

\begin{document}
\title{Coordinated Hybrid Precoding for Interference Exploitation in Heterogeneous Networks}

\author{Ganapati Hegde, ~\IEEEmembership{Student Member,~IEEE,}
    Christos Masouros,~\IEEEmembership{Senior Member,~IEEE,}
    Marius Pesavento,~\IEEEmembership{Member,~IEEE,}
    \thanks{G. Hegde and M. Pesavento are with the Communication Systems Group, Technische Universit{\"a}t Darmstadt, Darmstadt 64283, Germany. (e-mail: hegde@nt.tu-darmstadt.de; pesavento@nt.tu-darmstadt.de). C. Masouros is with the Department of Electronic \& Electrical Engineering, University College London, London WC1E7JE, U.K. (e-mail: c.masouros@ucl.ac.uk). The work of C. Masouros was supported by the Engineering and Physical Sciences Research Council Project EP/R007934/1.} 
    
    \thanks{ \textcopyright 2019 IEEE.  Personal use of this material is permitted.  Permission from IEEE must be obtained for all other uses, in any current or future media, including reprinting/republishing this material for advertising or promotional purposes, creating new collective works, for resale or redistribution to servers or lists, or reuse of any copyrighted component of this work in other works.	}
    
    \thanks{Digital Object Identifier 10.1109/LCOMM.2019.2933840}

}

\maketitle


\begin{abstract}
We consider a downlink multiuser massive MIMO system comprising multiple heterogeneous base stations with hybrid precoding architectures. To enhance the energy efficiency of the network, we propose a novel coordinated hybrid precoding technique, where the coordination between the base stations is aimed at exploiting interference as opposed to mitigating it as per conventional approaches. We formulate an optimization problem to compute the coordinated hybrid precoders that minimize the total transmit power while fulfilling the required quality of service at each user. Furthermore, we devise a low-complexity suboptimal precoding scheme to compute approximate solutions of the precoding problem. The simulation results reveal that the proposed coordinated hybrid precoding yields superior performance when compared to the conventional hybrid precoding schemes. 
\end{abstract}



\IEEEpeerreviewmaketitle

\vspace{-0.5cm}
\section{Introduction}

Network densification and massive MIMO are two major pillar technologies of 5G networks \cite{are_we_andrews, what_andrews, massive_hosseini}. In the network densification technique, each cell is deployed with multiple small base stations (BSs), e.g., micro BSs and pico BSs, along with the conventional macro BS \cite{survey_damnjanovic, optimal_hegde}. Due to the differences in features of the BSs such as the number of antenna elements and transmit power budget, such networks are called heterogeneous networks (HetNets). The BSs in the HetNets reuse the spectrum, which results in increased spectral efficiency and network throughput \cite{joint_fooladivanda}. A major challenge associated with the HetNets is the severe interference from a BS to users of the other BSs in a cell, as a consequence of the overlapping BS coverage areas \cite{massive_hosseini, enhanced_perez}. This inter-BS interference, if not carefully controlled, can drastically reduce the overall network capacity. To address this challenge, the coordinated scheduling is proposed in the literature, where the BSs communicate over the backhaul links and coordinate with each other to limit the interference \cite{coordinated_lee, centralized_ramos}. To further enhance the data rates, joint transmission with coordinated precoding is developed \cite{joint_zeng, joint_cheng}. In this technique, the precoding is performed jointly at multiple BSs to maximize the signal-to-interference-plus-noise ratio (SINR) at the users.

\begin{figure}[b!]
	\vspace{-0.65cm}
    \centering
    \includegraphics[scale=0.8]{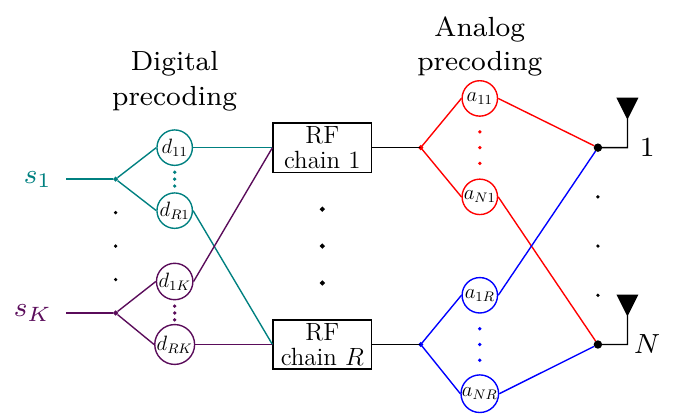}
    \caption{Hybrid analog-digital precoding architecture.}
    \label{fig_hybrid_precoding_architecture}
\end{figure}

\begin{figure}[t!]
    \centering
    \includegraphics[scale=0.2]{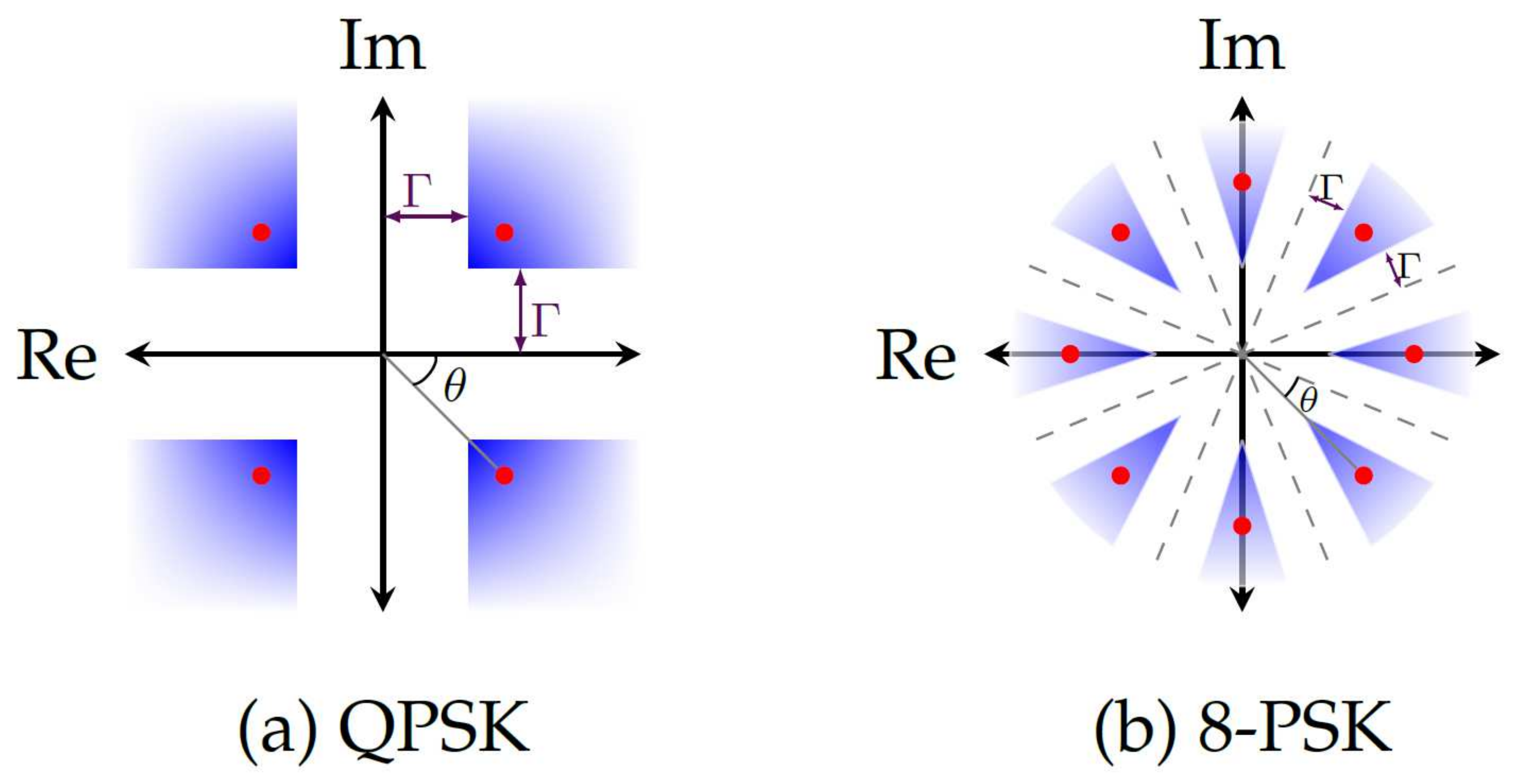}
    \caption{CI-regions (blue shaded area) of constellation symbols.}
    \label{fig_CI_region}
    \vspace{-0.5cm}
\end{figure}

In the massive MIMO systems, the BSs are equipped with hundreds of antennas \cite{massive_bjornson, massive_mimo_in_hoydis}. The degrees of freedom resulting from the large antenna array can be used to form spatially selective beams using the precoding technique \cite{an_overview_lu}. This technique enhances the signal powers at the intended users and reduces the interference powers at the unintended users. To reduce the hardware costs of massive MIMO systems, the hybrid analog-digital precoding architecture  \cite{spatially_ayach, beamforming_bogale, hybrid_beamforming_hegde,joint_hegde} is proposed, as shown in Fig.~\ref{fig_hybrid_precoding_architecture}. Unlike the conventional fully-digital precoding (FDP) \cite{iterative_schubert, optimal_bengtsson}, which requires a dedicated radio frequency (RF) chain for each antenna, the hybrid precoding requires fewer RF chains than the antenna elements. Even though hybrid precoding substantially reduces the hardware cost, inadvertently it also reduces the degrees of freedom for precoding and hence increases the transmit power when compared to the conventional FDP \cite{hybrid_li}. To overcome this shortcoming, power efficient constructive interference (CI)-based precoding \cite{exploiting_masouros, large_scale_amadori, constructive_khandaker, exploiting_timotheou, rethinking_zheng} has been extended to the hybrid precoding architecture \cite{analog_hegde, interference_hegde}. In the CI-based precoding, the knowledge of the channel state information and transmit symbols of all users is exploited to pre-equalize the transmit signals, such that the received signal at each user lies in the respective CI-region, that is the region in the constellation that is further away from the decision bounds. In this way, the interference inherent in the transmission is exploited constructively. A CI-region is separated from the corresponding decision boundaries by a \textit{threshold-margin} $\Gamma$, which controls the resulting quality of service (QoS), as illustrated in Fig.~\ref{fig_CI_region}.

The CI-precoding techniques in the literature do not trivially apply to the HetNet scenario. In this paper, we extend the CI-based hybrid precoding to the joint transmission in a HetNet in order to exploit the inter-BS interference for enhancing energy efficiency. We assume that the BSs in the network are connected to a central controller via low-latency backhaul links \cite{coordinated_saahashi, centralized_ramos, coordinated_lee}. The central controller gathers the spatial signatures characterizing the channels between the BSs and the users, and compute the hybrid precoders for all the BSs to serve the users jointly. An optimization problem is formulated to compute the CI-based hybrid precoders that minimize the total transmit power while achieving the required QoS at each user. Since the resulting problem is a nonconvex problem and difficult to solve optimally, we devise a suboptimal three-stage procedure to compute the coordinated hybrid precoders efficiently. As part of this procedure, mixed-integer linear problems are constructed to judiciously assign the RF chains to the users for continuous-valued and codebook-based analog precoding scenarios. Finally, the numerical results are provided to compare the performance of the proposed coordinated hybrid precoding with that of the conventional hybrid precoding schemes.

\vspace{-0.25cm}
\section{System Model}

We consider a multiuser downlink MIMO system comprising a set of $\mathcal{G} \bydef \{1,\ldots,G\}$ heterogeneous BSs. Each BS is equipped with multiple antenna elements in a fully-connected hybrid precoding architecture.  Let $N_g$ represent the number of antenna elements and $R_g$ indicate the number of RF chains at the $g$\myth BS, where $R_g \leq N_g$. The power budget at the $g$\myth BS is denoted by $P_g$. Let $\mathcal{K} \bydef \{1,\ldots, K\}$ denote a set of $K$ single antenna users in the network. The total number of RF chains from all the BSs, represented by $R_{\text{total}}$, is assumed to be larger than or equal to the number of users, i.e.,  $R_{\text{total}} \bydef \sum_{g \in \mathcal{G}} R_g \geq K$. Let $\mathcal{R} \bydef \{1,\ldots, R_{\text{total}}\}$ denote the set of all RF chains in the network. Moreover, $g_r \in \mathcal{G}$ is the index of the BS that contains the $r$\myth RF chain, $\forall r \in \mathcal{R}$. The transmit symbol intended for the $k$\myth user is denoted by $s_k$, which is drawn from uniformly distributed $M$-PSK symbols\footnote{We employ PSK-modulation here for notational simplicity. Nevertheless, the proposed techniques can be extended to other modulation formats following the principles in \cite{exploiting_li}.}. The threshold-margin at the $k$\myth user to achieve a requested QoS is denoted by $\Gamma_k$.

Let $\mathbf{A}_g \in \mathbb{C}^{N_g \times R_g}$ represent the analog precoding matrix at the $g$\myth BS comprising $R_g$ analog precoders of length $N_g$. Without loss of generality (w.l.o.g.), we assume that all PSs have an identical magnitude, which is denoted by $a$. The complex value of the PS that connects the {$r$\myth} RF chain to the {$n$\myth} antenna at the {$g$\myth} BS is denoted by $a_{gnr} = ae^{j\phi_{gnr}}$.  The vector $\mathbf{d}_{gk}$ indicates the digital precoder applied on symbol $s_k$ at the $g$\myth BS. The channel vector between the $g$\myth BS and the $k$\myth user is denoted by $\mathbf{h}_{gk} \in \mathbb{C}^{N_g \times 1}$. The received signal at the $k$\myth user from all the BSs is given by
\begin{align}
y_k = \sum_{g=1}^{G} \mathbf{h}_{gk}^{\trans} \mathbf{A}_g \sum_{m=1}^{K} \mathbf{d}_{gm}s_m + n_k, \quad \frall{k}, 
\end{align}
where $n_k$ stands for the additive white Gaussian noise at the $k$\myth user with zero-mean and variance $\sigma^2$.

\section{Proposed Solution}
In this paper, we aim to jointly compute the analog precoding matrix $\mathbf{A}_g$ and digital precoders $\mathbf{d}_{gk}, \frall{k}, \frall{g}$ that minimize the total transmit power while fulfilling the threshold-margin requirements of all users. An optimization problem to compute the corresponding precoders based on the CI-based precoding approach can be formulated as \cite{analog_hegde, interference_hegde}
\begin{subequations}
    \label{ch7_problem_coordinated_ori_hybrid_precoding}
    \begin{align}
    &\underset{ \{\mathbf{A}_g\}_{g \in \mathcal{G}}, \{\mathbf{d}_{gk}\}^{k \in \mathcal{K}}_{g \in \mathcal{G} } }{\operatorname{minimize\ }}  \sum_{g=1}^G\norm{ \mathbf{A}_g \sum_{m=1}^{K} \mathbf{d}_{gm} s_{m} }^2 \label{ch7_problem_coordinated_ori_hybrid_precoding_obj}\\
    &\operatorname{s.\ t.\ } \abs{\operatorname{Im}\left( s_{k}^{\conj}  \sum_{g=1}^{G} \ch{gk}^{\trans}  \mathbf{A}_g \sum_{m=1}^{K} \mathbf{d}_{gm} s_{m} \right)} \leq \nonumber \\ 
    & \hspace{1cm}\left( \operatorname{Re} \left(s_{k}^{\conj}  \sum_{g=1}^{G} \ch{gk}^{\trans} \mathbf{A}_g \sum_{m=1}^{K} \mathbf{d}_{gm} s_{m} \right) - \gamma_k \right) \tan \theta, \nonumber \\
    &\hspace{6cm} \forall k \in \mathcal{K}, \label{ch7_problem_coordinated_ori_hybrid_precoding_constraint_ci_region}\\
    &\hspace{0.82cm}\norm{ \mathbf{A}_g \sum_{m=1}^{K} \mathbf{d}_{gm} s_{m} }^2 \leq P_g, \quad \frall{g}, \label{ch7_problem_coordinated_ori_hybrid_precoding_power_budget} \\
    &\hspace{0.82cm}\abs{a_{gnr}} = a, \quad \frall{g}, \forall n \in \mathcal{N}_g, \forall r \in \mathcal{R}_g, \label{ch7_problem_coordinated_ori_hybrid_precoding_constraint_ap}
    \end{align}
\end{subequations}
where  $\theta \bydef \pi/M$,   $\gamma_k \bydef \Gamma_k/\sin\theta$, $\mathcal{N}_g \bydef \{1,\ldots, N_g\}$, and $\mathcal{R}_g \bydef \{1,\ldots, R_g\}$. In the above problem, the objective function (\ref{ch7_problem_coordinated_ori_hybrid_precoding_obj}) minimizes the total transmit power. The constraints in (\ref{ch7_problem_coordinated_ori_hybrid_precoding_constraint_ci_region}) force the nominal received signal at each user in the system to the respective CI-region\footnote{Due to the space constraints, the reader is referred to \cite{exploiting_masouros, interference_hegde} for the geometrical details of this formulation.}. The constraints in (\ref{ch7_problem_coordinated_ori_hybrid_precoding_power_budget}) limit the total transmit power at each BS to the respective power budget. The constraints in (\ref{ch7_problem_coordinated_ori_hybrid_precoding_constraint_ap}) administer the constant magnitude property of the analog precoding coefficients\footnote{In the codebook-based hybrid precoding, this constraint is replaced by a discrete constraint to select the analog precoders from the corresponding codebook.}.  By substituting  $\ech{gk} \bydef s_{k}^{\conj} \ch{gk} $ and treating the composite precoding term $\sum_{m = 1}^{K} \mathbf{d}_{gm} s_{m}$ as a single precoder $\mathbf{b}_{g}$, the problem (\ref{ch7_problem_coordinated_ori_hybrid_precoding}) can be reformulated as
\begin{subequations}
    \label{ch7_problem_coordinated_hybrid_precoding}
    \begin{align}
    &\underset{\{\mathbf{A}_g, \mathbf{b}_g\}_{g \in \mathcal{G}}}{\operatorname{minimize\ }} \sum_{g=1}^G \norm{\mathbf{A}_g \mathbf{b}_g}^2 \\
    &\text{s.\ t.\ } \abs{\operatorname{Im}\left(\sum_{g=1}^{G} \ech{gk}^{\trans} \mathbf{A}_g \mathbf{b}_g\right) } \leq \nonumber \\
    &\hspace{1cm} \left(\operatorname{Re}\left(\sum_{g=1}^{G}\ech{gk}^{\trans} \mathbf{A}_g \mathbf{b}_g\right) -  \gamma_k \right) \tan\theta,\ \forall k \in \mathcal{K}, \\
    &\hspace{0.67cm}\norm{\mathbf{A}_g \mathbf{b}_g}^2 \leq P_g, \quad \frall{g}, \\
    &\hspace{0.67cm}\abs{a_{gnr}} = a, \quad \frall{g}, \forall n \in \mathcal{N}_g, \forall r \in \mathcal{R}_g. \label{constraint_analog_precoder}
    \end{align}
\end{subequations}
The above problem is nonconvex and difficult solve due to the following reasons: 1)~The bilinear coupling between optimization variables $\mathbf{A}_g$ and  $\mathbf{b}_g$, and 2)~the nonconvex domain of the constant modulus variables $a_{gnr}$. To address this challenge, we propose to solve this problem suboptimally by performing the following tasks sequentially: 1)~The RF chain assignment, 2)~the analog precoding, and 3)~the digital precoding. 


\noindent\textbf{RF Chain Assignment:} In this stage, each RF chain in set $\mathcal{R}$ is assigned to a single user in set $\mathcal{K}$ such that each user in the network has at-least one dedicated RF chain. Here, we consider two scenarios based on the type of analog precoding employed at the BSs, namely, continuous-valued analog precoding, and codebook-based analog precoding.

\begin{itemize}
    \item \textbf{Continuous-Valued Analog Precoding:} 
    When the hybrid precoding architecture is equipped with the full-resolution PSs, the continuous-valued analog precoding is performed at the BS \cite{low_complexity_liang}. For the RF chain assignment in this scenario, firstly, we compute a channel gain matrix $\mathbf{Q} \in \mathbb{R}^{R_{\text{total}} \times K }$ that comprises the composite channel gains between the RF chains and the users. The composite channel gain between the $r$\myth RF chain and the $k$\myth user is defined as  $q_{rk} \bydef  \norm{\mathbf{h}_{g_rk}}^2$. Subsequently, we assign the RF chains to the users such that the total channel gain in the system is maximized while maintaining fairness among all users. The corresponding problem is formulated as a mixed-integer linear program (MILP) given by
    \begin{subequations}
        \label{ch7_problem_rf_assignment}
        \begin{align}
        \underset{\tau, \boldsymbol{\alpha}}{\operatorname{maximize\ }} & \sum_{r=1}^{R_{\text{total}}} \sum_{k=1}^{K}  \alpha_{rk} q_{rk} + \epsilon \tau\\
        \operatorname{s.\ t.\ } &\sum_{k=1}^{K} \alpha_{rk} \leq 1, \quad  \frall{r}, \label{ch7_problem_constraint_rf2single_user}\\
        &\sum_{r=1}^{R_{\text{total}}} \alpha_{rk} \geq 1, \quad \frall{k}, \label{ch7_problem_constraint_rf2each_user}\\
        & \sum_{r=1}^{R_{\text{total}}} \alpha_{rk} q_{rk} \geq \tau, \quad \frall{k}, \label{ch7_problem_constraint_total_channel_gain}\\
        & \alpha_{rk} \in \{0, 1\},\quad \frall{r}, \frall{k}.         
        \end{align}
    \end{subequations}
    In the above problem, the optimization matrix $\boldsymbol{\alpha}$ is a binary matrix of size $R_{\text{total}} \times K$, whose element $\alpha_{rk} = 1$ when the $r$\myth RF chain is assigned to the $k$\myth user, and $\alpha_{rk} = 0$ otherwise. The first term in the objective function represents the total channel gain of the system. The second term corresponds to the minimum sum channel gain among all users, which is denoted by $\tau$.  The constant $\epsilon$ is a scaling factor, which strikes a trade-off between the total channel gain and fairness among the users. The constraint (\ref{ch7_problem_constraint_rf2single_user}) confirms that an RF chain is not assigned to multiple users. The constraint (\ref{ch7_problem_constraint_rf2each_user}) enforces that at-least one RF chain is assigned to every user. The constraints in (\ref{ch7_problem_constraint_total_channel_gain}) administer fairness among the users.  The above problem can be solved using tools such as CVX (with integer-capable solves, e.g., Gurobi, MOSEK) \cite{cvx}.

    \item \textbf{Codebook-Based Analog Precoding:} 
    In this scenario, we assume that each BS comprises the codebook-based hybrid precoding architecture \cite{limited_alkhateeb, channel_alkhateeb}. Let $\boldsymbol{\mathcal{C}}_g$ indicate the codebook at the $g$\myth BS. We define a set $\boldsymbol{\mathcal{C}}  \bydef\boldsymbol{\mathcal{C}}_1 \cup \ldots \cup \boldsymbol{\mathcal{C}}_G$, which comprises codes from all the BSs. Let $C_\text{total} \bydef \card{\boldsymbol{\mathcal{C}}}$ denote the total number of codes in set $\boldsymbol{\mathcal{C}}$, and $\mathbf{C} \bydef [\mathbf{c}_1, \ldots, \mathbf{c}_{C_{\text{total}}}]$ be the codebook matrix. Moreover, the set $\mathcal{C} \bydef \{1,\ldots, C_{\text{total}}\}$ indicates the set of indices of codes in $\boldsymbol{\mathcal{C}}$, and the set $\mathcal{C}_g$ comprises the indices of codes of $\boldsymbol{\mathcal{C}}_g$ in set $\boldsymbol{\mathcal{C}}$. The index of the BS that contains the code $\mathbf{c}_c \in \boldsymbol{\mathcal{C}}$ is represented by $g_c$. 
    
    Now, we construct a channel gain matrix $\tilde{\mathbf{Q}} \in \mathbb{R}^{C_\text{total} \times K}$ comprising the channel gains between the codes in set $\boldsymbol{\mathcal{C}}$ and the users in set $\mathcal{K}$. The channel gain between the $c$\myth code belonging to the $g_c$\myth BS and the $k$\myth user is given by $\tilde{q}_{ck} \bydef \abs{\mathbf{c}_c^\trans \mathbf{h}_{g_c k}}^2$. The problem of assigning the codes to the users in order to maximize the total channel gain in the system while maintaining fairness among the users can be formulated as the following MILP. 
    \begin{subequations}
        \label{ch7_problem_rf_assignment_codebook}
        \begin{align}
        \underset{\tau, \boldsymbol{\alpha}}{\operatorname{maximize\ }} & \sum_{c=1}^{C_{\text{total}}} \sum_{k=1}^{K}  \alpha_{ck} \tilde{q}_{ck} + \epsilon \tau\\
        \operatorname{s.\ t.\ } &\sum_{k=1}^{K} \alpha_{ck} \leq 1, \quad  \frall{c}, \label{ch7_problem_codebook_constraint_rf2single_user}\\
        &\sum_{c=1}^{C_{\text{total}}} \alpha_{ck} \geq 1, \quad \frall{k}, \label{ch7_problem_codebook_constraint_rf2each_user}\\
        & \sum_{c=1}^{C_{\text{total}}} \alpha_{ck} \tilde{q}_{ck} \geq \tau, \quad \frall{k}, \label{ch7_problem_codebook_constraint_total_channel_gain}\\
        & \sum_{c \in \mathcal{C}_g} \sum_{k=1}^{K}\alpha_{ck} \leq R_g, \quad \frall{g}, \label{ch7_problem_codebook_constraint_total_users_per_BS}\\        
        & \alpha_{ck} \in \{0, 1\},\quad \frall{c}, \frall{k}.         
        \end{align}
    \end{subequations}
\end{itemize}  
In the above problem, the constraints in (\ref{ch7_problem_codebook_constraint_total_users_per_BS}) confirm that the number of codes selected from a BS is smaller than the number of RF chains available in that BS. The objective function and the other constraints play the same role as their counterparts in problem (\ref{ch7_problem_rf_assignment}). Similar to problem (\ref{ch7_problem_rf_assignment}), this problem can be solved using general purpose MILP solvers, e.g., Gurobi. 

\noindent\textbf{Analog Precoding:}
In the second stage, the analog precoding matrix is computed based on the RF chain assignment. In the continuous-valued analog precoding case, any suitable method, such as the CPC method \cite{analog_hegde}, can be employed to compute the analog precoders based on the channel vectors between the RF chains and their associated users. For example, if $\alpha_{rk} = 1$, then the elements of the analog precoder $\mathbf{a}_r$ are chosen according to $a_{g_r n r} = ae^{-\angle {h}_{g_r,k,n}}$. In the case of codebook-based analog precoding, the analog precoders at each BS are determined by the solution of problem (\ref{ch7_problem_rf_assignment_codebook}). For example, at the $g$\myth BS, the codes from $\boldsymbol{\mathcal{C}}_g$ for which $\alpha_{ck} = 1, \forall c \in \mathcal{C}_g, \frall{k}$ form the analog precoding matrix, i.e., $\hat{\mathbf{A}}_g \bydef \{\mathbf{c}_c\ \vert\ \alpha_{ck} =1, \forall c \in \mathcal{C}_g, \frall{k} \}$.

\noindent\textbf{Digital Precoding:}
For a given analog precoding matrix $\hat{\mathbf{A}}_g$, problem (\ref{ch7_problem_coordinated_hybrid_precoding}) can be readily reformulated as a convex optimization problem. The resulting problem is solved using e.g., CVX, to compute the digital precoder $\mathbf{b}_g$ at the $g$\myth BS, $\frall{g}$.

\section{Numerical Results}

For the simulation, we consider a HetNet comprising one macro BS and two identical pico BSs. The macro BS is equipped with $N_{\text{macro}} = 64$ transmit antennas, and the pico BSs are equipped with $N_{\text{pico}} = 32$ transmit antennas each. The number of RF chains at the macro BS $R_{\text{macro}}$ = 32 and at the pico BSs $R_{\text{pico}}$ = 16. There are $K = 64$ users deployed in the network. The path loss between the macro BS and users is modeled as $\text{PL}^{\text{macro}} (\text{dB}) = 128.1 + 37.6\times \log_{10} (d)$ and the path loss between a pico BS and users is modeled as $\text{PL}^{\text{pico}} (\text{dB}) = 140.7 + 36.7\times \log_{10}(d)$, where $d$ is the distance between a BS and a user in kilometers \cite{mobility_lopez}. Rayleigh small-scale fading is assumed. Noise power at each user is $\sigma^2=-60$ dBm. The optimization problems are solved by employing CVX with Gurobi solver \cite{cvx}. The results are averaged over 10,000 Monte Carlo runs.


\begin{figure}[t!]
    \centering
    \includegraphics[scale=0.7]{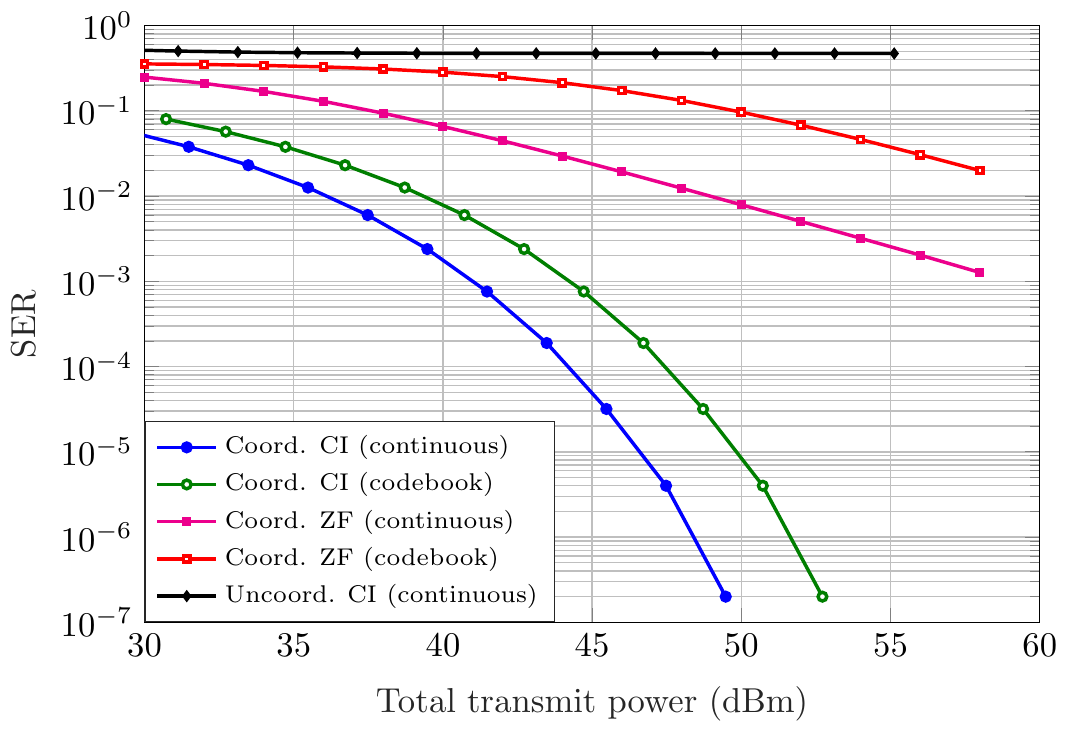}
    \caption{Performance comparison of the various hybrid precoding schemes for $N_{\text{macro}}= 64$, $N_{\text{pico}} = 32$, $R_{\text{macro}} = 32$,  $R_{\text{pico}} = 16$, $K=64$, and QPSK modulation.}
    \label{ch7_fig_ser_vs_p}
    \vspace{-0.5cm}
\end{figure}

\noindent\textbf{SER Performance:}
Figure~\ref{ch7_fig_ser_vs_p} plots SER vs. total transmit power (in dBm) achieved by the following schemes\footnote{For the coordinated CI-based methods, firstly, we compute the hybrid precoders for various TNR values. Subsequently, we calculate the total transmit power of the corresponding hybrid precoders. For the ZF-based methods, we fix the power budget in advance. Subsequently, we compute the hybrid precoders and numerically estimate the average SNRs. Finally, using the empirical relationship between SNR/TNR and SER given in Fig.~10 of \cite{interference_hegde} we compute the corresponding SERs for both CI-based and ZF-based methods.}: the proposed coordinated CI-based hybrid precoding with continuous-valued analog precoding (\textit{Coord. CI (continuous)}),  the proposed coordinated CI-based hybrid precoding with codebook-based analog precoding (\textit{Coord. CI (codebook)}), the coordinated schemes with continuous and codebook-based analog precoding followed by conventional interference suppression-based zero-forcing technique \textit{(Coord. ZF (continuous/codebook))} as digital precoding \cite{low_complexity_liang}, and the conventional uncoordinated CI-based hybrid precoding with continuous-valued analog precoding\footnote{In this scheme, each user is associated with the BS that provides the highest SINR value \cite{user_ye}. After the user association, each BS computes the CI-based hybrid precoders for its users independently \cite{analog_hegde, interference_hegde}.} \textit{(Uncoord. CI (continuous))}. The figure demonstrates that the coordinated CI-based hybrid precoding techniques result in significantly better performance when compared to the considered conventional hybrid precoding schemes. Due to the severe inter-BS interference the performance of the uncoordinated precoding scheme is extremely poor (with an increase in transmit power, even though useful signal power increases, the inter-BS interference increases as well). We also notice that the continuous-valued analog precoding, which requires expensive full-resolution PSs for the implementation, yields superior performance over the relatively-inexpensive codebook-based analog precoding in both CI-based and ZF-based methods.


\begin{figure}[t!]
    \centering
    \hspace{0cm}\includegraphics[scale=0.7]{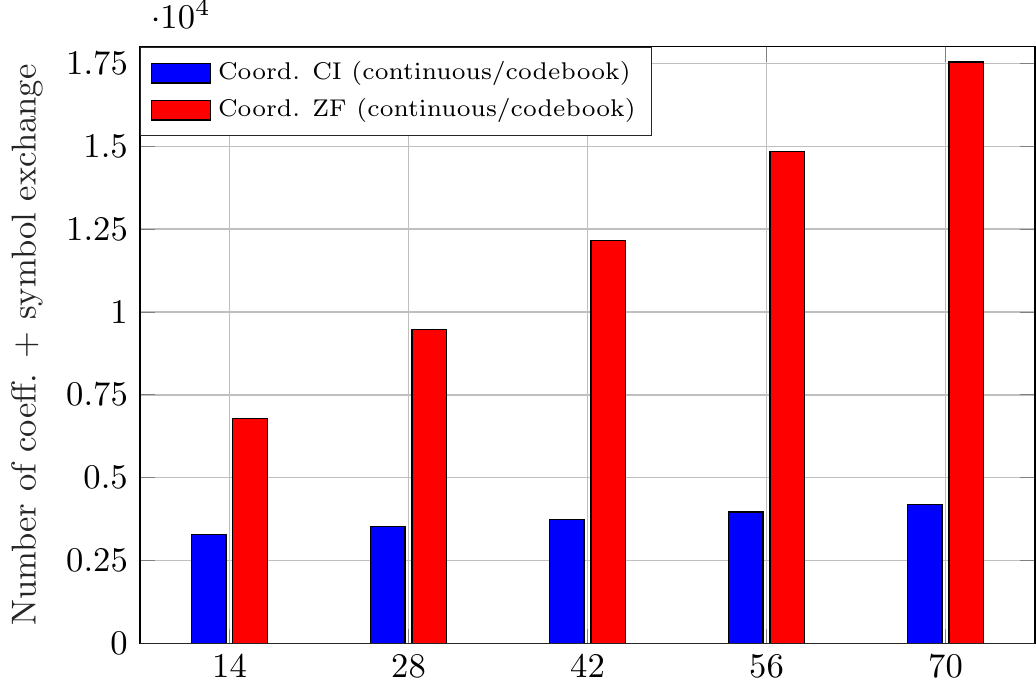}
    \caption{Amount of backhaul information exchange over the number of OFDM symbols (coherence time) for $N_{\text{macro}}= 64$, $N_{\text{pico}} = 32$, $R_{\text{macro}} = 32$,  $R_{\text{pico}} = 16$, and $K=64$.}
    \label{ch7_fig_backhaul_overhead}
    \vspace{-0.5cm}    
\end{figure}

\noindent\textbf{Coordination Overhead Analysis:} We quantify the coordination overhead incurred on network backhaul due to the coordinated precoding in terms of number of precoding coefficients and transmit symbols transmitted from the central controller to all BSs. Let $\Delta$ denote the number of OFDM symbol duration over which the channel is coherent. When the CI-based method employs an analog precoding scheme that is independent of the transmit symbols \cite{interference_hegde}, the number of analog precoding coefficients to be transmitted in the case of CI-based and ZF methods is $N_g\times R_g, \frall{g}$ over $\Delta$ symbol duration.
The CI-based method needs to transmit $R_g$ digital precoding coefficients (i.e., vector $\mathbf{b}_g$, which also constitutes the transmit symbols) to the $g$\myth BS for every symbol duration. In the case of ZF method, $R_g \times K$ digital precoding coefficients need to be sent once for every $\Delta$ symbol duration and $K$ symbols for every symbol duration to the $g$\myth BS, $\frall{g}$.  Accordingly, the total amount of information exchanged in the CI-based method is significantly smaller than that in the case of ZF method, as depicted in Figure \ref{ch7_fig_backhaul_overhead}.   

\vspace{-0.25cm}
\section{Conclusion}

In this paper, we proposed a coordinated CI-based hybrid precoding technique to serve the users jointly by all BSs in a HetNet. We formulated an optimization problem for computing the CI-based hybrid precoders with the smallest transmit power to fulfill the required QoS. To solve the problem efficiently, we devised a three-stage suboptimal scheme. In the first stage, the RF chains are assigned to the users. We formulated optimization problems for the RF chain assignment to maximize the total channel gain and maintain fairness. In the subsequent stages, the analog precoding and digital precoding are performed. The simulation results revealed the superior performance of the proposed scheme over the conventional hybrid precoding schemes.

\clearpage
\bibliographystyle{IEEEtran}
{\small \bibliography{main_source}}

\begin{thebibliography}{10}
\providecommand{\url}[1]{#1}
\csname url@samestyle\endcsname
\providecommand{\newblock}{\relax}
\providecommand{\bibinfo}[2]{#2}
\providecommand{\BIBentrySTDinterwordspacing}{\spaceskip=0pt\relax}
\providecommand{\BIBentryALTinterwordstretchfactor}{4}
\providecommand{\BIBentryALTinterwordspacing}{\spaceskip=\fontdimen2\font plus
\BIBentryALTinterwordstretchfactor\fontdimen3\font minus
  \fontdimen4\font\relax}
\providecommand{\BIBforeignlanguage}[2]{{%
\expandafter\ifx\csname l@#1\endcsname\relax
\typeout{** WARNING: IEEEtran.bst: No hyphenation pattern has been}%
\typeout{** loaded for the language `#1'. Using the pattern for}%
\typeout{** the default language instead.}%
\else
\language=\csname l@#1\endcsname
\fi
#2}}
\providecommand{\BIBdecl}{\relax}
\BIBdecl

\bibitem{are_we_andrews}
J.~G. {Andrews}, X.~{Zhang}, G.~D. {Durgin}, and A.~K. {Gupta}, ``Are we
  approaching the fundamental limits of wireless network densification?''
  \emph{{IEEE} Commun. Mag.}, vol.~54, no.~10, pp. 184--190, Oct. 2016.

\bibitem{what_andrews}
J.~G. Andrews, S.~Buzzi, W.~Choi, S.~V. Hanly, A.~Lozano, A.~C.~K. Soong, and
  J.~C. Zhang, ``What will {5G} be?'' \emph{{IEEE} J. Select. Areas Commun.},
  vol.~32, no.~6, pp. 1065--1082, Jun. 2014.

\bibitem{massive_hosseini}
K.~{Hosseini}, J.~{Hoydis}, S.~{ten Brink}, and M.~{Debbah}, ``Massive {MIMO}
  and small cells: {H}ow to densify heterogeneous networks,'' in \emph{Proc.
  {IEEE} Int. Conf. on Commun. {(ICC)}}, Jun. 2013, pp. 5442--5447.

\bibitem{survey_damnjanovic}
A.~Damnjanovic, J.~Montojo, Y.~Wei, T.~Ji, T.~Luo, M.~Vajapeyam, T.~Yoo,
  O.~Song, and D.~Malladi, ``A survey on {3GPP} heterogeneous networks,''
  \emph{{IEEE} Trans. Wireless Commun.}, vol.~18, no.~3, pp. 10--21, Jun. 2011.

\bibitem{optimal_hegde}
G.~Hegde, O.~D. Ramos-Cantor, Y.~Cheng, and M.~Pesavento, ``Optimal resource
  block allocation and muting in heterogeneous networks,'' in \emph{Proc.
  {IEEE} Int. Conf. on Acoustics, Speech and Signal Process. {(ICASSP)}},
  Shanghai, China, Mar. 2016, pp. 3581--3585.

\bibitem{joint_fooladivanda}
D.~{Fooladivanda} and C.~{Rosenberg}, ``Joint resource allocation and user
  association for heterogeneous wireless cellular networks,'' \emph{{IEEE}
  Trans. Wireless Commun.}, vol.~12, no.~1, pp. 248--257, Jan. 2013.

\bibitem{enhanced_perez}
D.~{Lopez-Perez}, I.~{Guvenc}, G.~{de la Roche}, M.~{Kountouris}, T.~Q.~S.
  {Quek}, and J.~{Zhang}, ``Enhanced intercell interference coordination
  challenges in heterogeneous networks,'' \emph{{IEEE} Wireless Commun.},
  vol.~18, no.~3, pp. 22--30, Jun. 2011.

\bibitem{coordinated_lee}
D.~{Lee}, H.~{Seo}, B.~{Clerckx}, E.~{Hardouin}, D.~{Mazzarese}, S.~{Nagata},
  and K.~{Sayana}, ``Coordinated multipoint transmission and reception in
  {LTE-Advanced}: {D}eployment scenarios and operational challenges,''
  \emph{{IEEE} Commun. Mag.}, vol.~50, no.~2, pp. 148--155, Feb. 2012.

\bibitem{centralized_ramos}
O.~D. Ramos-Cantor, J.~Belschner, G.~Hegde, and M.~Pesavento, ``Centralized
  coordinated scheduling in {LTE-Advanced} networks,'' \emph{{EURASIP} Journal
  on Wireless communications and Networking}, vol. 2017, no.~1, p. 122, Jul.
  2017.

\bibitem{joint_zeng}
Y.~{Zeng}, E.~{Gunawan}, Y.~L. {Guan}, and J.~{Liu}, ``Joint base station
  selection and linear precoding for cellular networks with multi-cell
  processing,'' in \emph{TENCON}, Nov. 2010, pp. 1976--1981.

\bibitem{joint_cheng}
Y.~{Cheng}, M.~{Pesavento}, and A.~{Philipp}, ``Joint network optimization and
  downlink beamforming for {CoMP} transmissions using mixed integer conic
  programming,'' \emph{{IEEE} Trans. Signal Process.}, vol.~61, no.~16, pp.
  3972--3987, Aug. 2013.

\bibitem{massive_bjornson}
E.~{Björnson}, E.~G. {Larsson}, and T.~L. {Marzetta}, ``Massive {MIMO}: {T}en
  myths and one critical question,'' \emph{{IEEE} Commun. Mag.}, vol.~54,
  no.~2, pp. 114--123, Feb. 2016.

\bibitem{massive_mimo_in_hoydis}
J.~Hoydis, S.~ten Brink, and M.~Debbah, ``Massive {MIMO} in the {UL/DL} of
  cellular networks: {H}ow many antennas do we need?'' \emph{{IEEE} J. Select.
  Areas Commun.}, vol.~31, no.~2, pp. 160--171, Feb. 2013.

\bibitem{an_overview_lu}
L.~Lu, G.~Y. Li, A.~L. Swindlehurst, A.~Ashikhmin, and R.~Zhang, ``An overview
  of massive {MIMO}: {B}enefits and challenges,'' \emph{{IEEE} J. Select.
  Topics in Signal Process.}, vol.~8, no.~5, pp. 742--758, Oct. 2014.

\bibitem{spatially_ayach}
O.~E. Ayach, S.~Rajagopal, S.~Abu-Surra, Z.~Pi, and R.~W. Heath, ``Spatially
  sparse precoding in millimeter wave {MIMO} systems,'' \emph{{IEEE} Trans.
  Wireless Commun.}, vol.~13, no.~3, pp. 1499--1513, Mar. 2014.

\bibitem{beamforming_bogale}
T.~E. Bogale and L.~B. Le, ``Beamforming for multiuser massive {MIMO systems}:
  {D}igital versus hybrid analog-digital,'' in \emph{Proc. {IEEE} Global
  Commun. Conf. {(GLOBECOM)}}, Austin, TX, USA, Dec. 2014, pp. 4066--4071.

\bibitem{hybrid_beamforming_hegde}
G.~Hegde, Y.~Cheng, and M.~Pesavento, ``Hybrid beamforming for large-scale
  {MIMO} systems using uplink-downlink duality,'' in \emph{Proc. {IEEE} Int.
  Conf. on Acoustics, Speech and Signal Process. {(ICASSP)}}, New Orleans, USA,
  Mar. 2017.

\bibitem{joint_hegde}
G.~Hegde and M.~Pesavento, ``Joint user selection and hybrid analog-digital
  beamforming in massive {MIMO} systems,'' in \emph{Proc. {IEEE} Sensor Array
  and Multichannel Signal Process. Workshop {(SAM)}}, Sheffield, UK, Jul. 2018.

\bibitem{iterative_schubert}
M.~Schubert and H.~Boche, ``Iterative multiuser uplink and downlink beamforming
  under {SINR} constraints,'' \emph{{IEEE} Trans. Signal Process.}, vol.~53,
  no.~7, pp. 2324--2334, Jul. 2005.

\bibitem{optimal_bengtsson}
M.~Bengtsson and B.~Ottersten, ``Optimal and suboptimal transmit beamforming,''
  \emph{Handbook of Antennas in Wireless Commun.}, Aug. 2001.

\bibitem{hybrid_li}
A.~{Li} and C.~{Masouros}, ``Hybrid analog-digital millimeter-wave {MU-MIMO}
  transmission with virtual path selection,'' \emph{{IEEE} Commun. Letters},
  vol.~21, no.~2, pp. 438--441, Feb. 2017.

\bibitem{exploiting_masouros}
C.~Masouros and G.~Zheng, ``Exploiting known interference as green signal power
  for downlink beamforming optimization,'' \emph{{IEEE} Trans. Signal
  Process.}, vol.~63, no.~14, pp. 3628--3640, Jul. 2015.

\bibitem{large_scale_amadori}
P.~V. Amadori and C.~Masouros, ``Large scale antenna selection and precoding
  for interference exploitation,'' \emph{{IEEE} Trans. Commun.}, vol.~65,
  no.~10, pp. 4529--4542, Oct. 2017.

\bibitem{constructive_khandaker}
M.~R.~A. Khandaker, C.~Masouros, and K.~K. Wong, ``Constructive interference
  based secure precoding: {A} new dimension in physical layer security,''
  \emph{{IEEE} Trans. on Inform. Forensics and Security}, vol.~13, no.~9, pp.
  2256--2268, Sep. 2018.

\bibitem{exploiting_timotheou}
S.~Timotheou, G.~Zheng, C.~Masouros, and I.~Krikidis, ``Exploiting constructive
  interference for simultaneous wireless information and power transfer in
  multiuser downlink systems,'' \emph{{IEEE} J. Select. Areas Commun.},
  vol.~34, no.~5, pp. 1772--1784, May 2016.

\bibitem{rethinking_zheng}
G.~Zheng \emph{et~al.}, ``Rethinking the role of interference in wireless
  networks,'' \emph{{IEEE} Commun. Mag.}, vol.~52, no.~11, pp. 152--158, Nov.
  2014.

\bibitem{analog_hegde}
G.~Hegde, C.~Masouros, and M.~Pesavento, ``Analog beamformer design for
  interference exploitation based hybrid beamforming,'' in \emph{Proc. {IEEE}
  Sensor Array and Multichannel Signal Process. Workshop {(SAM)}}, Sheffield,
  UK, Jul. 2018.

\bibitem{interference_hegde}
G.~{Hegde}, C.~{Masouros}, and M.~{Pesavento}, ``Interference
  exploitation-based hybrid precoding with robustness against phase errors,''
  \emph{{IEEE} Trans. Wireless Commun.}, vol.~18, no.~7, pp. 3683--3696, Jul.
  2019.

\bibitem{coordinated_saahashi}
M.~{Sawahashi}, Y.~{Kishiyama}, A.~{Morimoto}, D.~{Nishikawa}, and M.~{Tanno},
  ``Coordinated multipoint transmission/reception techniques for
  {LTE-Advanced},'' \emph{IEEE Wireless Communications}, vol.~17, no.~3, pp.
  26--34, Jun. 2010.

\bibitem{exploiting_li}
A.~Li and C.~Masouros, ``Exploiting constructive mutual coupling in {P2P}
  {MIMO} by analog-digital phase alignment,'' \emph{{IEEE} Trans. Wireless
  Commun.}, vol.~16, no.~3, pp. 1948--1962, Mar. 2017.

\bibitem{low_complexity_liang}
L.~Liang, W.~Xu, and X.~Dong, ``Low-complexity hybrid precoding in massive
  multiuser {MIMO} systems,'' \emph{{IEEE} Wireless Commun. Letters}, vol.~3,
  no.~6, pp. 653--656, Dec. 2014.

\bibitem{cvx}
M.~Grant and S.~Boyd, ``{CVX}: {M}atlab software for disciplined convex
  programming, version 2.1,'' \url{http://cvxr.com/cvx}, Mar. 2014.

\bibitem{limited_alkhateeb}
A.~Alkhateeb, G.~Leus, and R.~W. Heath, ``Limited feedback hybrid precoding for
  multi-user millimeter wave systems,'' \emph{{IEEE} Trans. Wireless Commun.},
  vol.~14, no.~11, pp. 6481--6494, Nov. 2015.

\bibitem{channel_alkhateeb}
A.~Alkhateeb, O.~E. Ayach, G.~Leus, and R.~W. Heath, ``Channel estimation and
  hybrid precoding for millimeter wave cellular systems,'' \emph{{IEEE} J.
  Select. Topics in Signal Process.}, vol.~8, no.~5, pp. 831--846, Oct. 2014.

\bibitem{mobility_lopez}
D.~{Lopez-Perez}, I.~{Guvenc}, and X.~{Chu}, ``Mobility management challenges
  in {3GPP} heterogeneous networks,'' \emph{{IEEE} Commun. Mag.}, vol.~50,
  no.~12, pp. 70--78, Dec. 2012.

\bibitem{user_ye}
Q.~{Ye}, B.~{Rong}, Y.~{Chen}, M.~{Al-Shalash}, C.~{Caramanis}, and J.~G.
  {Andrews}, ``User association for load balancing in heterogeneous cellular
  networks,'' \emph{{IEEE} Trans. Wireless Commun.}, vol.~12, no.~6, pp.
  2706--2716, Jun. 2013.

\end{thebibliography}


\end{document}